\documentclass[12pt,preprint]{aastex}

\begin{document}

\shortauthors{Frank, S. et.~al}

\shorttitle{Explanation of Disparate Absorption Statistics}

\title{Disparate \ion{Mg}{2} Absorption Statistics Towards Quasars and
Gamma-Ray Bursts:\\ A Possible Explanation}

\author{S.~Frank\altaffilmark{1},
	M.~C.~Bentz\altaffilmark{1},
	K.~Z.~Stanek\altaffilmark{1},
        S.~Mathur\altaffilmark{1},
	M.~Dietrich\altaffilmark{1},
	B.~M.~Peterson\altaffilmark{1},
        D.~W.~Atlee\altaffilmark{1}}

\altaffiltext{1}{\small Department of Astronomy, The Ohio State University, 
                 140 W. 18th Ave, Columbus, OH 43210}

\email{frank@astronomy.ohio-state.edu}

\begin{abstract}

We examine the recent report by \citeauthor{prochter06} that gamma-ray
burst (GRB) sight lines have a much higher incidence of strong
\ion{Mg}{2} absorption than quasar sight lines.  We propose that the discrepancy is due to the different beam sizes of GRBs and quasars, and that the intervening \ion{Mg}{2} systems are clumpy with the dense part of each cloudlet of a similar size as the quasars, i.e. $\lesssim 10^{16}\;$cm, but bigger than GRBs.
We also discuss observational predictions of our proposed model.  Most
notably, in some cases the intervening \ion{Mg}{2} absorbers in GRB
spectra should be seen varying, and quasars with smaller sizes should
show an increased rate of strong \ion{Mg}{2} absorbers.  In fact, our
prediction of variable \ion{Mg}{2} lines in the GRB spectra has been now
confirmed by \citeauthor{hao07}, who observed intervening \ion{Fe}{2} and
\ion{Mg}{2} lines at $z=1.48$ to be strongly variable in the multi-epoch
spectra of $z=4.05$ GRB\,060206.

\end{abstract}


\section{Introduction}

As some of the most luminous objects in the Universe, gamma-ray bursts
(GRBs) have been established as important cosmological probes.  In
particular, the transient optical afterglows of GRBs are important for
absorption line studies along lines of sight that are not associated
with quasars (cf. \citealt{vreeswijk03,chen05}).

Recently, \citeauthor{prochter06} (2006, hereafter P06) reported
statistically significant evidence for a higher incidence of strong
\ion{Mg}{2} $\lambda \lambda 2796,2803$ absorbers towards GRB sight
lines than towards quasars of comparable redshifts. More precisely,
they found 15 strong (rest frame equivalent width $W_{\rm r} > 1.0$
\AA) \ion{Mg}{2} absorbing systems in a sample of 12 GRBs with optical
follow-up spectroscopy allowing for both \ion{Mg}{2} doublet
components to be identified at $> 3 \sigma$ confidence level. Based on
the study of a Sloan Digital Sky Survey (SDSS) quasar sample with over
50,000 quasars containing more than 7,000 \ion{Mg}{2} systems with
$W_{\rm r} > 1.0$\,\AA, P06 expect only 3.4 such systems in quasar
spectra over the redshift path interval of the GRB sample.  Thus,
about four times as many strong \ion{Mg}{2} absorbers are observed
along the line of sight to GRBs compared to quasar sight lines.

P06 propose and quickly dismiss three effects to explain the discrepancy
in the observed and expected incidence of \ion{Mg}{2} absorbers along
GRB sight lines.  First, they suggest that dust associated with the
\ion{Mg}{2} absorbers could obscure faint quasars resulting in a lower
number of observations of such systems along quasar sight
lines. Secondly, the absorbing gas could simply be associated with
the GRB event, which they dismiss as unphysical because of the
relativistic speeds necessary and the intrinsically narrow absorption
line widths. As a third possibility, P06 propose that the GRB population
could be gravitationally lensed by these absorbers; this does not seem
to be the case as the lensing magnification necessary would be evident
by bright foreground galaxies or multiple images of the source.  P06
therefore conclude that ``at least one of the fundamental beliefs on
absorption line research must be flawed.''

In this {\it Letter}, we propose a geometric solution for the observed difference in incidences and explore observational signatures of the proposed
solution.

\section{Geometric Differences Between GRBs and Quasars}

Given that the P06 result is not affected by selection effects,
gravitational lensing or \ion{Mg}{2} absorbers local to GRB hosts, we
are left with the scenario that GRB sight lines trace the same
intervening absorber population as quasars. These intervening
\ion{Mg}{2} absorbers thus ``know'' nothing about the background light
sources, and yet their number densities are different in GRB and quasar
sightlines. Here we suggest a simple geometric explanation for the
different number of strong \ion{Mg}{2} absorbers per unit redshift along
sight lines towards quasars and GRBs. We postulate that (1) the GRB
sizes are, on {\em average}, smaller than quasar sizes, and (2) the
\ion{Mg}{2} absorbers are clumpy (or have density structure), with
characteristic sizes on the order of GRB beam sizes. These simple
assumptions lead to the observed differences due to dilution of the
\ion{Mg}{2} column density as seen by the larger quasar beams.  We
elaborate on this suggestion below. 

In order to demonstrate the effect of beam size on dilution of observed
equivalent widths, we have constructed a simple ``toy'' model. Assume that the \ion{Mg}{2} clouds have a clumpy density structure:

\begin{eqnarray}
\rho(r) = \left\{
\begin{array}{ll}
\rho_{0}{} & ;\  r\leq r_{0},
\\
\rho_{0}(r/r_0)^{-k} & ;\  r > r_{0}
\end{array}
\right.
\end{eqnarray}

\begin{figure}
	\includegraphics[angle=270, width=\columnwidth]{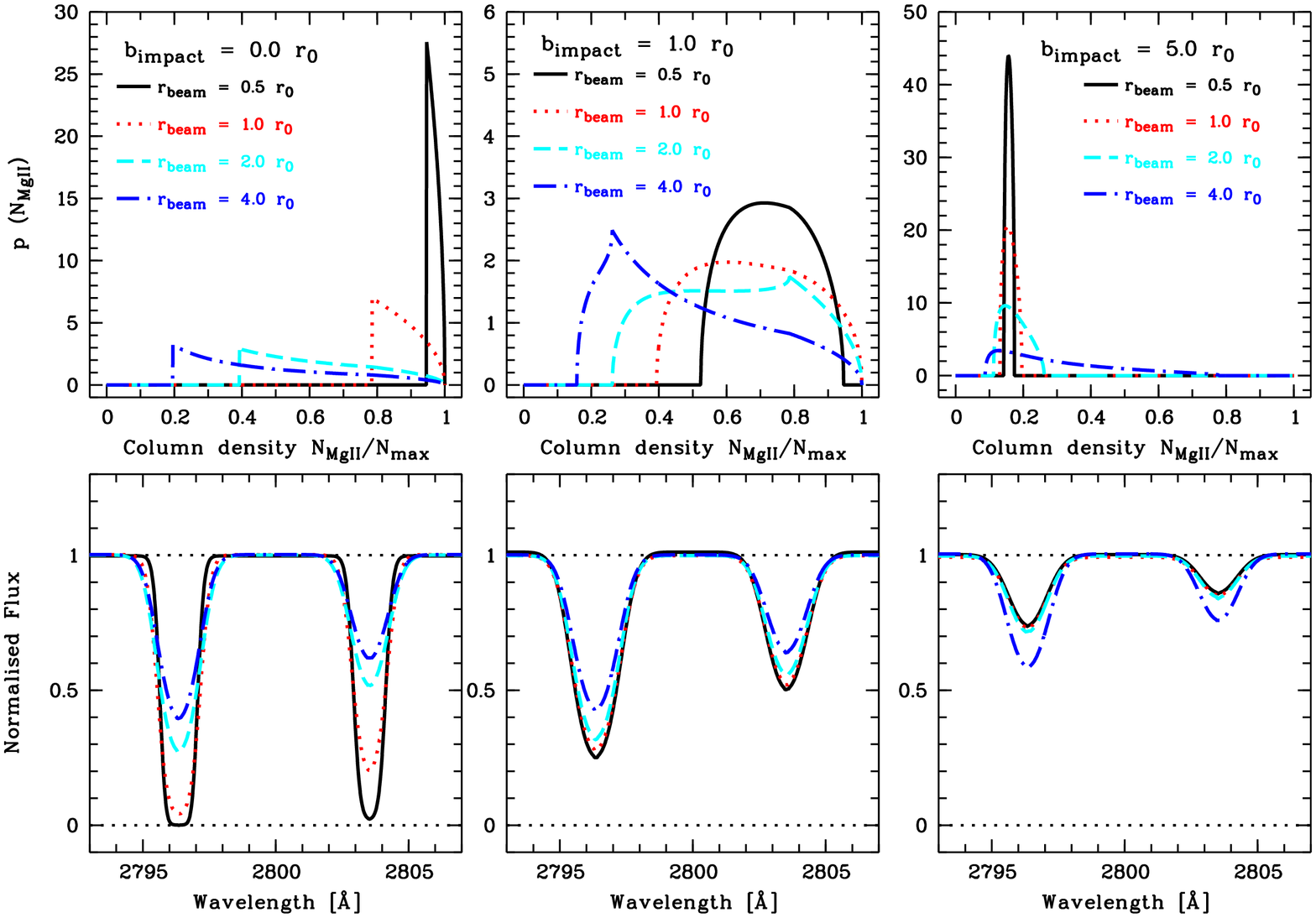} 	
	\caption{{\bf Upper panels :} The (normalised) column density distribution functions for beams of different size and impact parameters relative to the core size of the absorbers $r_{0}$. These functions have been calculated for the specific case of k=3 for the density profile of the absorber. Note how the average column density for low impact parameters (left panel) is a strongly varying function of the beam size. For beams just grazing the dense part of the absorbers (middle panel), the distributions of column densities sampled become more similar, and finally for large impact parameters (right panel), smaller beams sample even lower density regions than their large counterparts. {\bf Lower panels :} Theoretical spectra computed from the column density distributions above. For the specific parameters (maximum column density N$_{max}$, and Doppler parameter $b$) of the individual Voigt profiles superposed here, see the text. Note the overall trend to lower equivalent widths from low to high impact parameters (left to right), and how the pronounced differences of the profile depths in the direct hit case (left panel) almost vanish for higher impact parameters (middle and right panels). }
\label{profiles_and_spectra}
\end{figure}

\begin{figure}
	\includegraphics[angle=270, width=\columnwidth]{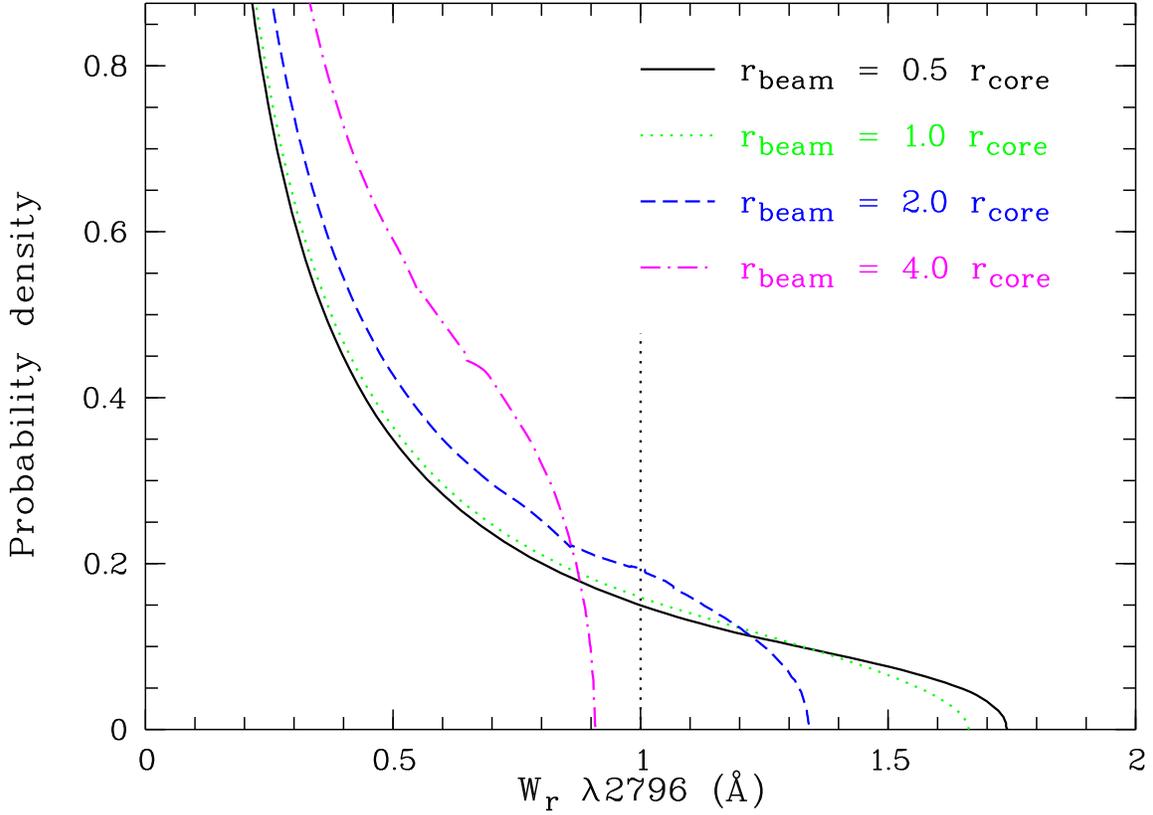}
	\caption{The probability density of obtaining $W_{\rm
	r}$(2796\AA) for beams of various sizes under the assumption of
	random locations of the cloud core relative to the beams. The
	transformation between the sampled average column density and
	$W_{\rm r}$ was chosen such that an absorber sampling 0.4
	N$_{max}${} yields a line of $W_{\rm r}$ = 1.0 \AA. For details
	see text. The vertical dotted line indicates the boundary
	between strong and weak absorbers used by \citet{prochter06}{}
	for their statistical analysis. Note how the larger beams can
	only reach low $W_{\rm r}$s, readily explaining the scarcity of
	such absorbers found in quasar studies. The calculations
	presented here are for the specific case of k=3.01 for the
	power-law density coefficient.}
\label{equivalent_width}
\end{figure}

Thus, the density of an individual, spherically symmetric absorbing
cloud is constant, $\rho_{0}$, within a certain core radius r$_0$,
outside of which it drops like a power-law.  The maximum column density
of a single beam hitting the cloud core is N$_{MgII} = (2+\frac{2}{k-1})
\rho_{0} r_{0}$. On the other hand, if a single beam does not intersect
the core, but passes through the cloud at an impact parameter $b$, then
it samples a column density N$_{MgII} \propto b^{-k+1}$.  We calculate
the difference in the \ion{Mg}{2} column densities produced by such a
cloud, absorbing randomly located, uniform brightness beams of different
sizes. The column density profile sampled by the beam depends on the size of the beam relative to the core size of the absorber, and how far away from the cloud core the beam passes the absorber, i.e. the impact parameter $b$. Fig. \ref{profiles_and_spectra}{} shows various examples for different scenarios for the specific case of k=3.01 for the powerlaw-slope of the absorber density. In the upper left panel, the beams hits the absorber centrally. Small beams in this case sample much higher column densities than beams larger than the core. In the middle panel, where the centre of the beams just grazes the absorber core, the differences in the column densities seen by the background source become less prominent, but bigger beams still have on average lower average column densities.  This situation is reversed for cases where the impact parameter is large enough so that no part of the beams reaches the dense core of the absorber. Not only does the average column density for each beam drop quickly, but now larger beams can sample higher densities due to some parts being much closer to the denser parts of the absorber.\\
We model the \ion{Mg}{2}{} absorber spectra, superposing Voigt profiles of different column densities $N_{MgII} / N_{max}${} and weighing factors $p(N_{MgII})${} taken from the column density distribution functions of figure \ref{profiles_and_spectra}. The lower panels of figure \ref{profiles_and_spectra}{} shows the examples computed from the scenarios in the upper panels. The parameters for these profiles were chosen such that a beam sampling an average column density of 0.4 N$_{max} = 0.4 \times 2.995\; \rho_{0} r_{0}${} results in a line of 1.0 \AA{} for a Doppler parameter of $b$ = 80 km/s\footnote{This value of the Doppler parameter reflects roughly both the instrumental resolution of the SDDS spectra, used by P06 in their analysis, and the typical velocity spread of \ion{Mg}{2}{} absorber systems (see e.g. \citet[]{charlton1998}). Individual components of such systems have much lower $b${} values (Churchill 2000).}. Assuming a core radius r$_{0} = 5 \times 10^{15}\;$cm (cf. Hao et al. 2007), leads to an estimate for the density
of such a cloudlet of $\rho_{0}$(\ion{Mg}{2})$\sim 5\times10^{-3}$
cm$^{-3}$. As we stress below, the exact parameters of this ``toy
model'' are not relevant, but are designed to show how geometric
``dilution'' of observed equivalent width works.\\        
The lower left panel of figure \ref{profiles_and_spectra}{} shows how different beam sizes and impact parameters affect the strength of the \ion{Mg}{2}{} absorbers. The differences in the equivalent widths for small and large beams are especially pronounced for impact parameters less  than $b_{imp} \sim 0.5 \times r_{0}$(lower left panel). Notice how slowly moving out to larger $b_{imp}${} primarily affects the small beams, while the structure of the absorber for the larger beams remains almost unchanged (lower middle panel). At large distances, the reversal in column density sampling mentioned above, finds it equivalent in the larger beams now producing stronger absorbers (lower right panel).\\
By way of calculating such spectra, we can transform the column density distribution into the observable rest-frame equivalent width $W_{\rm r}$.
Fig. \ref{equivalent_width}{} shows the probability density of obtaining $W_{\rm r}$(2796\AA) for different beam sizes, again assuming the specific case of k=3.01 for the power-law density coefficient. This probability density is obtained by
integrating the impact parameter distribution out to some fiducial value
b$_{max}${} under the assumptions that all impact parameters have the
same likelihood, and that there are no other nearby cores within
b$_{max}$. As we stress below, the exact parameters of this ``toy
model'' are not relevant, but are designed to show how geometric
``dilution'' of observed equivalent width works. As shown in figure 2,
we observe a range of $W_{\rm r}$ with small beams, but with bigger
beams only smaller values of $W_{\rm r}$ are observed due to dilution
effects.

Thus, the higher incidence of GRB sight-lines with strong
\ion{Mg}{2}{} absorbers reported by \citet{prochter06} follows
naturally from the beam dilution for the larger quasar beam. What we
have discussed above is a ``toy model'' designed to explain this
observational result. Determination of exact beam sizes of quasars or
GRBs, exact core size distribution of \ion{Mg}{2}{} ``cloudlets'' or
their radial density profile is not the intent here. We note,
nonetheless, that invoking somewhat different beam sizes and geometry
distributions, as well as absorber core size distributions in connection
with spatial clustering of multiple such ``cloudlets'' would change the
details of the $W_{\rm r}$ distribution, but would not change our main
results substantially.

This simple models explains one more observational result: Various
quasar absorption line studies have revealed that the number of
\ion{Mg}{2} absorbers strongly declines with $W_{\rm r}$. \citet{ppb06}
find $f(W_{\rm r})= 490.37 W_{\rm r}^{-2.245}$ for $W_{\rm r} > 1$\,\AA,
where $f(W_{\rm r})$ is the number of systems per unit equivalent width
for their SDSS Data Release 3 quasar sample. In our model such a decline
follows automatically as producing such systems becomes harder for the larger QSO beams since they always sample regions of low column densities. Again, we point out that reproducing the exact shape of the equivalent width distribution is not our goal here.

\section{Discussion}
   
The differences in the \ion{Mg}{2} absorption properties of quasars and GRBs
can be easily understood if the continuum emitting regions of
quasars are several times bigger than those of GRBs.  Let us examine
whether this assumption is reasonable.

In the standard thin disk model \citep{shakura73}, the source size for a
quasar with an AB magnitude $m_{AB}$ and an absorber at observed
wavelength $\lambda$ is

\begin{equation}
R \simeq \left( 7 \times 10^{15} \right) \left( { c D_A \over H_0 } \right) 
\left( {\lambda \over \mu\hbox{m}} \right)^{3/2} 
10^{-0.2(m_{AB}-19)} h^{-1} \hbox{cm}.
\end{equation}

On the other hand, the optical afterglow ring radius of a GRB evolves as
(\citealt{waxman97, loeb98})
\begin{equation}
R_s(t)= 4.1\times 10^{15} \left({E_{52}\over n_1}\right)^{1\over 8} 
(1+z_{\rm s})^{-5/8} (t/\rm hour)^{5/8}~{\rm cm} 
\end{equation}
\noindent where the factor involving the source redshift $z_{\rm s}$
is due to the cosmic time dilation, $E_{52}$ is the
``isotropic-equivalent'' of the energy release in units of $10^{52}~{\rm
erg~s^{-1}}$ and $n_1$ is the ambient gas density in units of ~$1~{\rm
cm^{-3}}$.  For the well-observed GRB\,060206 at $z_{\rm GRB}=4.05$, the
radius of the afterglow ring is estimated to be $5\times 10^{15}$~cm
\citep{hao07} a few hours after the burst, i.e., when their spectra were
obtained (see discussion below).

Initial estimates thus imply the GRB emitting regions to be comparable
to those of quasars. In reality, however, the GRB size can be
significantly smaller than given by the simplistic formula above, if the
optical emission from a GRB is dominated by small sections of the
potentially visible ring of emission.  In many cases the light curve of
a GRB afterglow is observed to be very irregular, with significant
($\sim 20\%$) deviations from a power-law decay on timescales of 20-30
minutes approximately 1~day after the burst (e.g., GRB 021004:
\citealt{bersier03}).  \citet{nakar03} proposed that such irregular
light curves could result from a patchy shell model of the GRB jet,
i.e. the jet structure would have strong angular dependence. This would
effectively result in a smaller size of the GRB afterglow compared to
the numbers above (see also \citealt{ioka05}) --- only a part of the
narrow ring would dominate the brightness, reducing the size of the GRB
beam by a factor of $\sim 10$ down to few $\times \sim 10^{14}\,$cm. We
note that the light curve of GRB\,060206, for which Hao et al. (2007)
have observed strongly variable intervening \ion{Mg}{2} and \ion{Fe}{2}
lines, has a very irregular optical light curve (Stanek et
al. 2007). 

What do we know about the sizes of AGNs and GRBs from direct
observational measurements? An intensive multiwavelength monitoring
program of NGC~7469 yielded a time delay of 1.5 light days between
variations in the optical continuum measured at 6962\,\AA\
\citep{collier98} relative to those in the ultraviolet continuum
measured at 1315\,\AA\ \citep{wanders97}.  This is consistent with the
report of \citet{sergeev05}, where they find evidence for lag times of
$\sim 1-20$ light days for the $V$, $R$, and $I$ bands relative to the
$B$ band in 14 nearby ($z < 0.1$) Seyferts.  This would imply continuum
sizes of order $10^{15}$--$10^{16}$~cm for nearby quasars and sizes that
are an order of magnitude or so larger for the more distant, more
massive quasars observed in the \ion{Mg}{2} studies.  On the other hand,
the observation of microlensing of lensed quasars is generally less
compatible with such large source sizes.  The systematic measurement of
quasar disk sizes using microlensing is just beginning, but three recent
results lead to sizes of $10^{15.0-15.5}$, $10^{14.0-14.5}$, and
$10^{15.5-16}h^{-1}$~cm for the gravitational lenses HE0435--1223
\citep{kochanek06}, SDSS0924+0219 \citep{morgan06} and Q2237+0305
\citep{kochanek04}, respectively, and these scales are broadly
consistent with the expectations from thin disk theory.  However, other
recent quasar microlensing studies claim that optical emission regions
of lensed quasars could be much larger than expected from basic disk
models by factors of $\sim$10-100 (Pooley et al. 2006; Hawkins
2006). Finally, we note that the incidence of \ion{Mg}{2} absorbers
along blazar sight lines is also enhanced relative to those of more
typical quasars \citep{stocke97}. If, once again, difference in beam
sizes is the cause of this effect, then quasar sizes are bigger than
blazar sizes. \citet{kataoka00} estimate a blazar beam size of 0.01~pc,
right in the range of sizes discussed above and so quasar sizes must be
larger than $3\times 10^{16}$cm. The only observational constraints on
GRB emission region sizes come from the radio scintillations observed in
GRB\,970508 \citep{waxman98}. These observations do not constrain the
GRB sizes well, but find that they are broadly consistent with the
theoretical expectation. To summarize, our key assumption of quasar
optical emission regions being on average several times larger than GRB
emission regions is allowed by current constraints on their sizes, both
theoretical and from direct observations.

The other key ingredient of our suggestion is that the \ion{Mg}{2}
absorbers have to be patchy with a characteristic size of $\sim
10^{16}\;$cm.  While no direct size measurement of \ion{Mg}{2}
``clouds'' exists, we have some information that is not inconsistent
with our proposed, small size.  \citet{rauch02} studied
\ion{Mg}{2} absorbers in the three lines of sight towards the lensed
quasar Q$2237+0305$ (Huchra's lens) and found that the individual
lines of sight do not show the same \ion{Mg}{2} cloudlets (see for
example their striking Fig.~10, showing a \ion{Mg}{2} system at
$z=0.827$). So, the \ion{Mg}{2} absorbers are clearly patchy on scales
of $<200-300\;$pc.  \citet[]{ding03}, analyzing strong \ion{Mg}{2}{}
absorbers towards PG1634+706, find evidence for structures on scales
of parsecs to hundreds of parsecs with densities similar to what we
derive for our model, and temperatures of $\sim$10,000
K. Interestingly, they point out that there are hints for
\ion{Mg}{1} pockets of much higher density which could be as small
as $\sim$0.001 pc, and conclude that ``they are
pervasive in the disk of the Milky Way'' and should therefore be seen
in absorption through other galaxies. High resolution studies towards
globular clusters and binary stars have revealed substantial
structure within the Galactic disk ISM down to scales as low as
proposed here (e.g. \citealt{andrews01}{} and references therein).

It is also of interest that Fischera et al. (2004,2005), analysing the effect of turbulence-induced clumpiness, arrive at similar column density distribution functions (cf. e.g. Fischera et al. (2004) Fig. 5 with our Figure 1, upper panel). In their model, a physical mechanism (turbulence) creates a clumpy structure in the large-scale structure of a dust screen of galactic scale. Such a 'clumpy screen' then can create the observed difference in the extinction law of dust, and attenuation law seen in the continuum light of galaxies. 

We stress again that we are proposing a new idea in which geometric
effects account for the differences in absorption incidence in quasar
and GRB sightlines. This is not meant to be a rigorous model in which
exact sizes of quasars, GRBs or \ion{Mg}{2} clouds are calculated. We
show, however, that the assumed parameters are reasonable, though the
observational or theoretical constraints are not strong. If future
observations show that GRB sizes are not smaller than quasars or that
\ion{Mg}{2} are much bigger, it will clearly invalidate our proposed
solution.

 Our ``toy model'' leads to specific observational predictions:\\
1. GRBs are, by nature, dynamic objects.  As a result, our
explanation necessarily predicts that the strength and structure of the
absorption lines along many GRB sight lines should vary over time as the
GRB ring evolves and the beam size changes. This indeed has now been
observed in at least one case by \citet{hao07} who discovered strongly
variable intervening \ion{Mg}{2} and \ion{Fe}{2} lines at $z=1.48$ in
the line of sight to the $z=4.05$ GRB\,060206. \\
2. Blazars are also extremely variable objects
(cf. \citealt{stein76,angel80}), and thus the same effect should be seen
along blazar sight lines. \\
3. We expect the number of weak absorption systems observed in the
spectra of GRB to be lower than that for equivalent quasar
populations. This follows logically from a number conservation
argument. This prediction has not been tested yet due to the
observational difficulty of detecting weak features in the transient GRB
phenomena and the still rather small number of GRB afterglow spectra. \\
4. Additionally, as the size of the quasar continuum-emitting
region is dependent on the luminosity, a careful study of different
luminosity populations of quasars should also reveal a difference in
the incidence and strength of intervening absorbers. \\
5. Micro-lensing of strongly lensed
quasars can probe the $\sim 100$~AU scale intergalactic \ion{Mg}{2}
cloudlets, as proposed by \citet{dong06}. In short, \ion{Mg}{2} lines
seen in the spectra of lensed quasars should be seen evolving.

To summarise, we propose that the difference is
due to the larger beam size of quasars relative to GRBs, and the
comparable size of \ion{Mg}{2} clouds.  This leads to specific
observational predictions, of which we discuss several. One of our
predictions has already been confirmed by \citet{hao07}. If
further confirmed, it will lead to improved constraints on the sizes of
quasars, GRBs and \ion{Mg}{2} absorbers, as well as confirming our
proposed explanation for the puzzling result of P06.

\acknowledgments 

We are grateful for useful suggestions by Sara Ellison and Pat
Hall.   We also 
thank the participants of the morning ``Astronomy Coffee'' at the
Department of Astronomy, The Ohio State University, for the daily and
lively astro-ph discussion, one of which prompted us to investigate
the problem described in this paper.  MCB is supported by a Graduate
Fellowship of the National Science Foundation.

%\bibliographystyle{apj}
%\bibliography{refs}

\clearpage

\end{document}